# B → ρ Decay Form Factors from Covariant Confined Quark Model

Aidos Issadykov [1,2,3]

[1] *Joint Institute for Nuclear Research, 6 Joliot-Curie, 141980 Dubna, Moscow region, Russia.*
[2] *Institute of Nuclear Physics, Ministry of Energy, 1 Ibragimova, 050032 Almaty, Republic of Kazakhstan.*
[3] *Al-Farabi Kazakh National University, 71 al-Farabi, 050038 Almaty, Republic of Kazakhstan.*

*Corresponding author: issadykov@jinr.ru*

**Abstract.** We evaluate $B \to \rho$ transition form factors in the full kinematical region within the covariant confined quark model. We compare the obtained results with the results from light-cone sum rules. The calculated form factors can be used to calculate the $B^+ \to \rho^+ \mu^+ \mu^-$ rare decay branching ratio.

## INTRODUCTION

The $b \to s\ell^+\ell^-$ and $b \to d\ell^+\ell^-$ processes are forbidden at tree-level in Standart Model and sensitive to New Physics contributions in loops. The $b \to d$ transition is suppressed than more $b \to s$ due to CKM matrix elements. However it is interesting to study decays proceeds via flavour-changing neutral-current (FCNC) transition. The $b \to d$ transition decays observed by LHCb collaboration for $B^+ \to \pi^+ \mu^+ \mu^-$ [1, 2] and $B_s^0 \to K^{*0} \mu^+ \mu^-$ [3] decays.

A search for the rare decay $B^0 \to \rho^0(770)\ (\to \pi^+\pi^-)\mu^+\mu^-$ governed by $b \to d$ weak transition is described in LHCb collaboration paper[4]. The $B \to \rho$ transition form factors were studied in light-cone sum rules [5, 6]. In view of these development, we calculate $B \to \rho$ form factors within the covariant confined quark model (CCQM).

## MODEL

The covariant confined quark model [7] is an effective quantum field approach to hadronic interactions based on an interaction Lagrangian of hadrons interacting with their constituent quarks. The value of the coupling constant follows form the compositeness condition $Z_H = 0$, where $Z_H$ is the wave function renormalization constant of the hadron. Matrix elements of the physical processes are generated by a set of quark loop diagrams according to the $1/N_c$ expansion. The ultraviolet divergences of the quark loops are regularized by including vertex functions for the hadron-quark vertices. These functions also describe finite size effects related to the non-pointlike hadrons. The quark confinement [8] is built-in through an infrared cutoff on the upper limit of the scale integration to avoid the appearance of singularities in matrix elements. The infrared cutoff parameter $\lambda$ is universal for all processes. The covariant confined quark model has limited number of parameters: the light and heavy constituent quark masses, the size parameters which describe the size of the distribution of the constituent quarks inside the hadron and the infrared cutoff parameter $\lambda$. They are determined by a fit to available experimental data. We fix $\Lambda$ parameters according to the experimental value of leptonic decay constants [9].

In calculations we used next values of the model parameters which are shown in Eq. (1).

| $m_{u/d}$ | $m_s$ | $m_c$ | $m_b$ | $\lambda$ | $\Lambda_B$ | $\Lambda_\rho$ | $m_B$ | $m_\rho$ | |
|---|---|---|---|---|---|---|---|---|---|
| 0.241 | 0.428 | 1.67 | 4.68 | 0.181 | 1.963 | 0.624 | 5.279 | 0.775 | GeV |

(1)

More details concerning the model can be found in our previous papers [10,11].
Below, we list the definitions of the dimensionless invariant transition form factors together with the covariant quark model expressions that allow one to calculate them. We closely follow the notation used in previous papers [10, 12].

$$\langle V(p_2,\epsilon_2)_{[\bar{q}_3 q_2]}|\bar{q}_2 O^\mu q_1|P_{[\bar{q}_3 q_1]}(p_1)\rangle =$$
$$= \frac{\epsilon_\nu^\dagger}{m_1+m_2}\left(-g^{\mu\nu}P\cdot q A_0(q^2) + P^\mu P^\nu A_+(q^2) + q^\mu P^\nu A_-(q^2) + i\varepsilon^{\mu\nu\alpha\beta}P_\alpha q_\beta V(q^2)\right), \quad (2)$$

$$\langle V(p_2,\epsilon_2)_{[\bar{q}_3 q_2]}\left|\bar{q}_2\left(\sigma^{\mu\nu}q_\nu(1+\gamma^5)\right)q_1\right|P_{[\bar{q}_3 q_1]}(p_1)\rangle =$$
$$= \epsilon_\nu^\dagger\left(-(g^{\mu\nu}-q^\mu q^\nu/q^2)P\cdot q a_0(q^2) + (P^\mu P^\nu - q^\mu P^\nu P\cdot q/q^2)a_+(q^2) + i\varepsilon^{\mu\nu\alpha\beta}P_\alpha q_\beta g(q^2)\right). \quad (3)$$

We use $P = p_1 + p_2$, $q = p_1 - p_2$ and the on–shell conditions $\epsilon_2^\dagger \cdot p_2 = 0$, $p_i^2 = m_i^2$, $m_1 = m_B$ and $m_2 = m_\rho$. Since there are three quark species involved in the transition, we have introduced a two–subscript notation $w_{ij} = m_{q_j}/(m_{q_i} + m_{q_j})$ $(i,j = 1,2,3)$ such that $w_{ij} + w_{ji} = 1$. The form factors defined in Eq. (3) satisfy the physical requirement $a_0(0) = a_+(0)$, which ensures that no kinematic singularity appears in the matrix element at $q^2 = 0$.

## NUMERICAL RESULTS

The form factors are calculated in the full kinematical region of momentum transfer squared and results of our numerical calculations are with high accuracy approximated by the parametrization

$$F(q^2) = \frac{F(0)}{1-as+bs^2}, \quad s = \frac{q^2}{m_1^2}. \quad (4)$$

the relative error is less than 1%. The values of $F(0)$, $a$ and $b$ are listed in Table 1.

**TABLE I**. Parameters for the approximated form factors in Eq. 4

| | $A_0$ | $A_+$ | $A_-$ | $V$ | $a_0$ | $a_+$ | $g$ |
|---|---|---|---|---|---|---|---|
| $F(0)$ | 0.33 | 0.23 | -0.24 | 0.26 | 0.24 | 0.24 | 0.23 |
| $a$ | -0.51 | -1.36 | 0.05 | -1.49 | -0.02 | -0.05 | -0.05 |
| $b$ | -0.31 | 0.37 | -0.0005 | 0.47 | -0.0003 | 0.0005 | 0.0006 |

The curves are depicted in Figure 1.

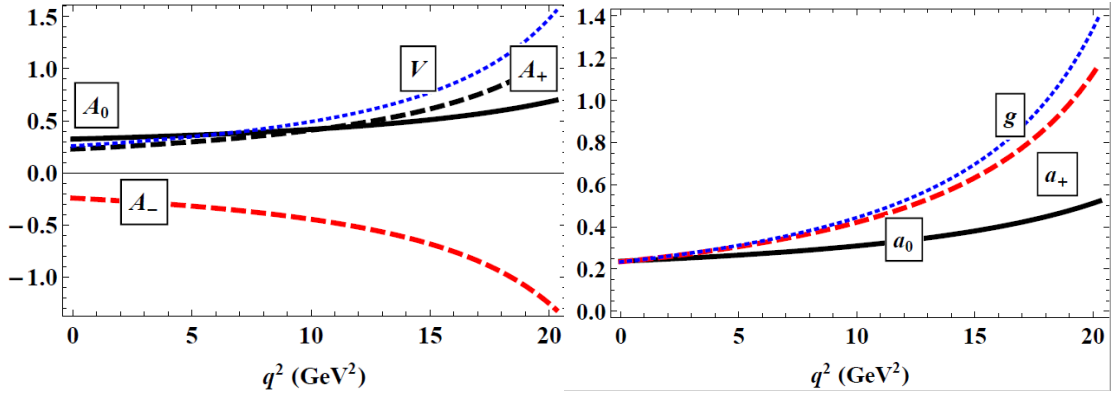

**FIGURE 1**. The $q^2$-dependence of the vector and axial form factors (left plot) and tensor form factors (lower plot) for $B \to \rho$ decay

The obtained errors of the fitted parameters are of the order of 10%. Indeed it follows that form factors at $q^2 = 0$ were calculated with 10% uncertainties. This implies at least 10% uncertainty in form factors in the full kinematical region of momentum transfer squared.

For reference it is useful to relate the above form factors to those used, e.g., in Ref. [5] (we denote them by the superscript $^C$). The relations read

$$A_0 = \frac{(m_1 + m_2)}{(m_1 - m_2)} A_1^c, \qquad A_+ = A_2^c,$$

$$A_- = \frac{2m_2(m_1 + m_2)}{q^2}(A_3^c - A_0^c), \qquad V = V^c,$$

$$a_0 = T_2^c, \qquad g = T_1^c, \qquad a_+ = T_2^c + \frac{q^2}{m_1^2 - m_2^2} T_3^c. \tag{5}$$

We note in addition that the form factors (5) satisfy the constraints

$$A_0^c(0) = A_3^c(0)$$
$$2m_2 A_3^c(q^2) = (m_1 + m_2) A_1^c(q^2) - (m_1 - m_2) A_2^c(q^2). \tag{6}$$

Since $a_0(0) = a_+(0) = g(0)$ we display in Table 2 the form factors $A_0^c(0) = (m_1 - m_2)[A_0(0) - A_+(0)]/(2m_2)$, $A_1^c(0) = A_0(0)(m_1 - m_2)/(m_1 + m_2)$, $A_2^c(0) = A_+(0)$, $T_1^c(0) = g(0)$, $T_3^c(0) = \lim_{q^2 \to 0}(m_1^2 - m_2^2)(a_+ - a_0)/q^2$ obtained in our model and compare them with those from light-cone sum rule [5,6].

**TABLE II**. The form factors of $B \to \rho$ transition at maximum recoil $q^2 = 0$ in the covariant confined quark model (CCQM) in comparison with the values of other works and approaches.

|      | $V^c(0)$ | $A_0^c(0)$ | $A_1^c(0)$ | $A_2^c(0)$ | $T_1^c(0)$ | $T_3^c(0)$ |
|------|----------|------------|------------|------------|------------|------------|
| CCQM | 0.26+0.03 | 0.29±0.03 | 025±0.03 | 0.23±0.02 | 0.23±0.02 | 0.19 ±0.02 |
| [5]  | 0. 323 | 0.303 | 0.242 | 0.221 | 0.267 | 0.176 |
| [6]  | 0.327±0.031 | 0.356±0.042 | 0.262±0.026 | 0.297±0.035 | 0.272±0.026 | 0.747±0.076 |

## DISCUSSION AND CONCLUSION

In this paper we present form factors in the full kinematical region for B→ρ transition within the covariant confined quark model (CCQM). The obtained results in a good agreement with the results from light-cone sum rules within errors. The next step will be the investigating of the branching ratio for the rare $B^+ \to \rho^+ \mu^+ \mu^-$ decay and compare it with available data.

## ACKNOWLEDGMENTS

Author A. Issadykov is grateful for the support by the JINR, grant number 19-302-03.